\documentclass{ws-ijmpa}

\begin{document}

%

\def\nocropmarks{\vskip5pt\phantom{cropmarks}}

\let\trimmarks\nocropmarks      

%

\markboth{Jen-Chieh Peng}
{Status of Spin Physics - Experimental Summary}

%
\catchline{}{}{}
%

\setcounter{page}{1}

\title{Status of Spin Physics - Experimental Summary}

\author{\footnotesize Jen-Chieh Peng}

\address{Physics Division, Los Alamos National Laboratory, 
Los Alamos, New Mexico, 87545, U.S.A.\\
and\\
Department of Physics, University of Illinois,
Urbana, Illinois, 61801, U.S.A.
}

\maketitle

\begin{abstract}
The current status of spin physics experiments, based
on talks presented at the Third Circum-Pan-Pacific Symposium on
High Energy Spin Physics held in Beijing, 2001, is summarized
in this article. Highlights of recent experimental results at 
SLAC, JLab, and DESY, as well as future plans at these facilities 
and at RHIC-spin are discussed.
\end{abstract}

\section{Introduction}	

The purpose of this article is to summarize the experimental status
of spin physics based on the material presented at the
Third Circum-Pan-Pacific Symposium on High Energy Spin Physics.
A total of 12 experimental talks covering 
recent results and future plans at DESY, JLab, SLAC, and
RHIC-spin were presented at this Symposium. These talks
successfully convey the sense of excitement in this field
through the presentation of many interesting recent experimental
results, as well as exciting prospects for future experiments.

Several excellent review articles on high-energy spin physics
have been published 
recently\cite{filippone01,bunce00,jaffe01,anselmino01,jakob01}.
Many new experimental 
results were presented at this Symposium for the first time, 
indicating that spin physics has become one of the most active 
areas of research in nuclear and particle physics.

The physics topics covered by the 12 experimental talks can be
grouped into the following categories:

\begin{itemize}

\item Polarized structure functions

$g_1(x,Q^2)$, $g_2(x,Q^2)$, $h(x,Q^2)$, their integrals, and the GDH sum
rule.

\item Quark and gluon helicity distributions

$\Delta q(x,Q^2)$ and $\Delta G(x,Q^2)$.

\item Transversity distributions

$\delta q(x,Q^2)$

\item Generalized parton distributions

Deeply virtual Compton scattering (DVCS).

\item Other related topics

Longitudinal spin transfer in $\Lambda$ production,
exclusive meson productions, $A_1^n$ at large $x$, etc.

\end{itemize}

In the following sections, we will discuss the recent progress in
and the future prospects for these various areas of researches.

\section{Polarized Structure Functions}

\subsection{$g_1(x,Q^2)$ and $\Gamma_1$}

\begin{figure}[t]
\vspace{65mm}
\centering{\includegraphics{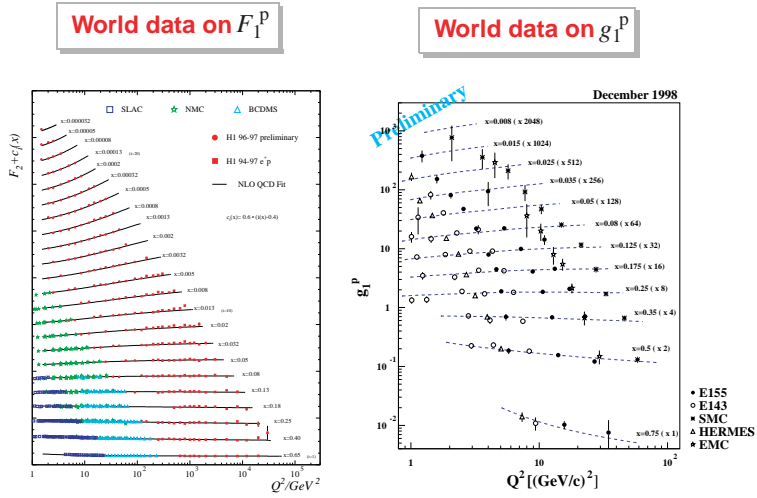}}
\caption{Collection of $F_1^p(x,Q^2)$ and $g_1^p(x,Q^2)$ data 
taken from Refs. 3,6.}
\label{fig:f1pscaling}
\end{figure}

Following the discovery of the ``spin-crisis" in the
late '80s, extensive efforts have been devoted to accurate measurements of 
the spin-dependent structure functions $g_1^p(x,Q^2)$ and 
$g_1^n(x,Q^2)$.  A series of experiments at SLAC (E142, E143, 
E154, E155, E155x), at CERN (EMC, SMC), and at DESY (HERMES), 
have measured $g_1^p$ and $g_1^n$ over a broad range of 
$x$ and $Q^2$. Scaling violation of $g_1^p$ is now clearly observed, 
as shown in Fig.~\ref{fig:f1pscaling}. The $Q^2$ evolution 
of $g_1^p(x,Q^2)$ is strikingly similar to that of the 
spin-averaged structure function
$F_1^p(x,Q^2)$. In particular, the E155 collaboration 
recently found\cite{E155_00} that
the ratio $g_1(x,Q^2)/F_1(x,Q^2)$ of all existing data
can be parameterized as

\begin{equation}
g_1/F_1 = x^\alpha (a + bx + cx^2) (1 + \beta / Q^2).
\label{Eq:g1f1ratio}
\end{equation}

\noindent The values of $\beta$, $-0.04 \pm 0.06~(0.13 \pm 0.45)$
for the proton (neutron), are consistent with zero and indicate
that $g_1$ and $F_1$ have very similar $Q^2$ dependences.

The extensive data on $g_1(x,Q^2)$ allow accurate determinations
of the integrals 
$\Gamma_1^{p,n}(Q^2) = \int_{0}^{1} g_1^{p,n}(x,Q^2)dx$ for the
proton and the neutron,
as well as $\Gamma_1^p(Q^2) - \Gamma_1^n(Q^2)$.
Table 1 lists the results of these integrals from 
recent NLO analysis of existing data by the E155\cite{E155_00} 
and SMC\cite{SMC_98,SMC_98a} collaborations.
While the values of $\Gamma_1^p$ and $\Gamma_1^n$ are different from
the predictions of Ellis and Jaffe\cite{ellis74} who assumed SU(3)
flavor symmetry and an unpolarized strange sea, the data are
in good agreement with the prediction of
the Bjorken sum rule\cite{bjorken66}.

\begin{table}[htbp]
\ttbl{30pc}{$\Gamma_1^p$, $\Gamma_1^n$, and $\Gamma_1^p - \Gamma_1^n$
from recent NLO analysis.}
{\begin{tabular}{cccccc}\\
\multicolumn{6}{c}{} \\[6pt]\hline
$Q^2(GeV^2)$ & $\Gamma_1^p$ & $\Gamma_1^n$ &$\Gamma_1^p - \Gamma_1^n$ 
&$\Gamma_1^p - \Gamma_1^n$ (theory) &Ref.\\ \hline
5 & $0.118\pm 0.008$ & $-0.058\pm0.009$ & $0.176\pm0.007$ & $0.182\pm0.005$ &
7 \\
10 & $0.120\pm 0.016$ & $-0.078\pm0.021$ & $0.198\pm0.023$ & $0.186\pm0.023$ &
8,9 \\
5 & $0.121\pm 0.018$ & $-0.075\pm0.021$ & $0.174+0.024-0.012$ 
& $0.181\pm0.003$ & 9 \\ \hline
\end{tabular}}
\end{table}

As the Bjorken sum rule is now quite well tested, it is not surprising that
the experimental activities on $g_1(x)$ are winding down. Nevertheless, 
there are other interesting aspects of $g_1(x)$ worthy of further 
studies. First, the behavior of $g_1(x)$ at low $x$, $x< 0.003$, is
not yet known. Perturbative QCD calculations, based on fits to existing 
data, give predictions for $g_1(x)$ at low $x$ very different from 
Regge and other models\cite{close94,kwiecinski99}. The largest uncertainty
on the $\Gamma_1^p$ determination also comes from the unmeasured low-$x$
region. Future polarized $e-p$ collider is required for exploring the
low-$x$ region\cite{hughes99}. Another interesting topics is the behavior
of $g_1(x)$ integral at low $Q^2$, to be discussed next.

\subsection{$\Gamma_1(Q^2)$ at low $Q^2$ and the generalized GDH integral}

How does $\Gamma_1(Q^2)$ evolve as $Q^2 \to 0$? This question 
is closely related to the Gerasimov-Drell-Hearn (GDH) sum 
rule\cite{gerasimov66,drell66}:

\begin{equation}
\int_{\nu_0}^{\infty} [\sigma_{1/2}(\nu) - \sigma_{3/2}(\nu)] \frac{d\nu}{\nu}
= -\frac{2\pi^2\alpha}{M^2} \kappa^2.
\label{Eq:GDH}
\end{equation}

\noindent The GDH sum rule, based 
on general physics principles (causality, unitarity,
Lorentz and gauge invariances) and dispersion relation, relates the total
absorption cross sections of circularly polarized photons on longitudinally
polarized nucleons to the static properties of the nucleons. 
In Eq.~\ref{Eq:GDH}, $\sigma_{1/2}$ and $\sigma_{3/2}$ are the photo-nucleon
absorption cross sections of total helicity of $1/2$ and $3/2$, $\nu$ is
the photon energy and $\nu_0$ is the pion production threshold, $M$ is
the nucleon mass and $\kappa$ is the nucleon anomalous magnetic moment.
The GDH sum rule predictions are -205 $\mu$b and -233 $\mu$b for the
proton ($\kappa_p = +1.793$) and neutron ($\kappa_n = -1.913$), respectively.

A first measurement\cite{ahrens01} of the helicity dependence of 
photoabsorption cross section on the proton was recently carried out at 
MAMI (Mainz) and the contribution to the GDH sum 
was found\cite{ahrens01} to be
$-226\pm5\pm12$ $\mu$b for the photon energy range $200 < \nu < 800$ MeV.
Using the Unitary Isobar model\cite{drechsel99} and a Regge 
model\cite{bianchi99} to estimate the contributions from unmeasured 
energy regions, the integral is found\cite{ahrens01} 
to be $-210\mu$b, consistent with 
the GDH sum rule. Measurements at higher energies are either underway 
or being prepared at ELSA, JLab, and SLAC.

The GDH integral in Eq.~\ref{Eq:GDH} can be generalized from real photon
absorption to virtual photon absorption with non-zero $Q^2$:

\begin{eqnarray}
I_{GDH}(Q^2) & \equiv & \int_{\nu_0}^{\infty} [\sigma_{1/2}(\nu,Q^2) 
- \sigma_{3/2}(\nu,Q^2)] \frac{d\nu}{\nu} \nonumber \\
 & = & \frac{8\pi^2\alpha}{M} 
\int_{0}^{x_0} \frac{g_1(x,Q^2) - \gamma^2g_2(x,Q^2)}{K} \frac{dx}{x},
\label{Eq:GGDH}
\end{eqnarray}

\noindent where $K=\nu \sqrt{1+\gamma^2}$ is the flux factor of the virtual
photon, $\gamma^2 = Q^2/\nu^2$ and $x_0 = Q^2/2M\nu_0$. The 
generalized GDH integral connects the 
helicity structures of the nucleons measured in high-energy
electron DIS to those in low-energy photo-absorption at the resonance region.
Assuming the validity of the Burkhardt - Cottingham
sum rule\cite{burkhardt70}, $\int_{0}^{1}g_2(x,Q^2)dx = 0$, 
it follows that for $\gamma \to 0$ 
Eq.~\ref{Eq:GGDH} becomes

\begin{equation}
I_{GDH}(Q^2) = \frac{16\pi^2\alpha}{Q^2} \Gamma_1(Q^2). 
\label{Eq:GGDH1}
\end{equation}

\noindent Eq.~\ref{Eq:GGDH1} shows that 
the $Q^2$-dependence of the generalized GDH integral
is directly related to the $Q^2$-dependence of $\Gamma_1$. 
$\Gamma_1^p$ is known to be
positive at high $Q^2$ and the GDH sum rule (Eq.~\ref{Eq:GDH})
predicts $\Gamma_1^p = 0$ 
at $Q^2=0$ with a negative slope for $d\Gamma_1^p(Q^2)/dQ^2$, therefore,
$\Gamma_1^p(Q^2)$ must become negative at
low $Q^2$.

\begin{figure}[htbp]
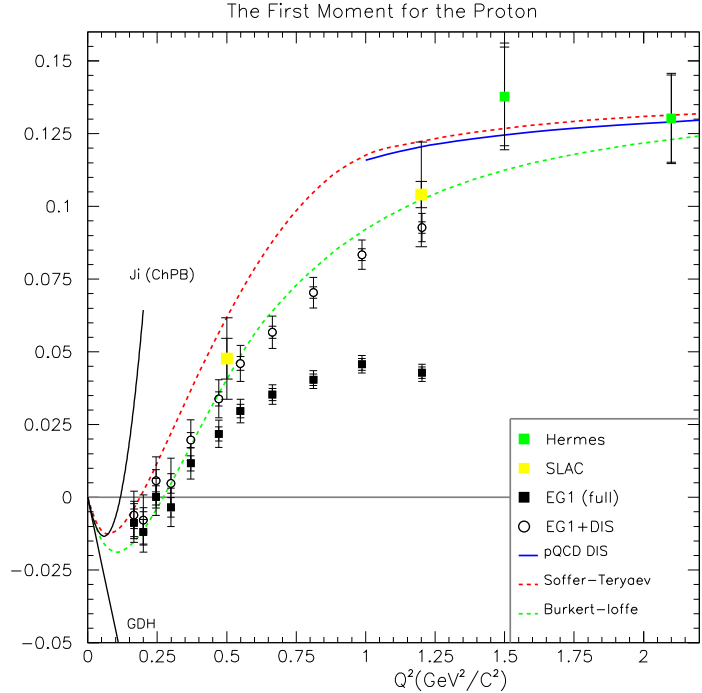
 
\PSFIG{peng2}{4.0in}{0}
\caption{\small Preliminary result of $\Gamma_1^p(Q^2)$ from 
CLAS$^{21}$.
Data from SLAC and HERMES are also shown. Theoretical curves
are from Refs. 22,23.}
\label{fig:gammap}
\end{figure}

The GDH integrals at low $Q^2$ have recently been measured 
in several JLab experiments. As reported by Griffioen\cite{griffioen01} 
in this Symposium, inclusive double spin asymmetry have been measured
at Hall-B using the CLAS spectrometer with polarized 
electron beams of 2.5 and 4.2 GeV scattered off longitudinally 
polarized $N\vec H_3$ and $N\vec D_3$ targets.
The kinematical region covered in this experiment corresponds to
$0.1 < Q^2 < 2.7$ GeV$^2$ and $W < 2.5$ GeV. Preliminary results on
$\Gamma_1^p(Q^2)$ extracted from this experiment are shown in 
Fig.~\ref{fig:gammap}. These data indeed show that $\Gamma_1^p$ changes
sign around $Q^2 = 0.3$ GeV$^2$. The origin of the sign-change 
can be attributed to the competition between $\Delta(1232)$ and
higher nucleon resonances. At the lowest $Q^2$, the $\Delta(1232)$ has a
dominant negative contribution to $\Gamma_1^p$. However, at larger
$Q^2$, higher mass nucleon resonances take over to have a net positive
$\Gamma_1^p$. As shown in Fig.~\ref{fig:gammap}, the strong $Q^2$ 
dependence of $\Gamma_1^p(Q^2)$ is well reproduced by the calculation
of Burkert and Ioffe\cite{burkert92}.

\begin{figure}[htbp]
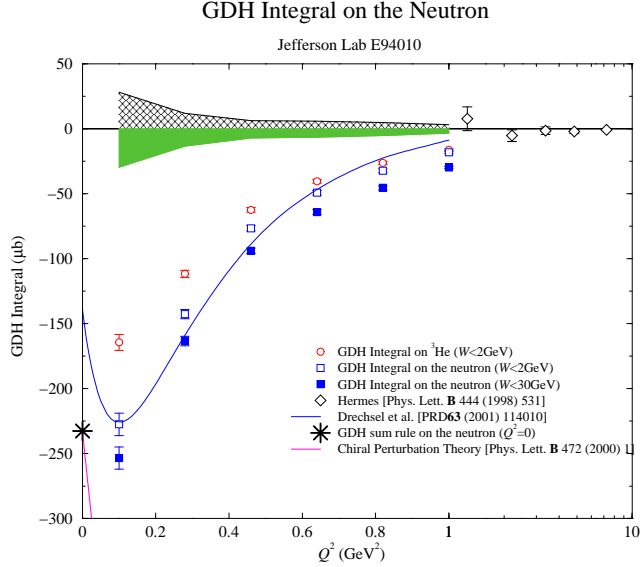
 
\PSFIG{peng3}{3 in}{-90}
\caption{Near final results of the JLab 
Hall-A measurement$^{24}$
of the Generalized GDH sum for neutron and $^3He$.}
\label{fig:nGDH}
\end{figure}

As reported by Chen\cite{chen01} in this Symposium, an extensive spin
physics program has been underway using the JLab Hall-A spectrometers.
In particular, neutron spin-dependent structure functions, $g_1^n(x,Q^2)$ 
and $g_2^n(x,Q^2)$, have been measured using an intense polarized
electron beam on either
longitudinally or transversely polarized $^3He$ target.
Preliminary results\cite{chen01} on the generalized 
GDH integral for neutron and $^3He$ are shown in Fig.~\ref{fig:nGDH}.
In contrast to the proton case, the strong negative 
contribution to the GDH integral from the 
$\Delta(1232)$ resonance now dominates the entire measured $Q^2$ range. 
The data appear to approach the GDH sum rule value at $Q^2 = 0$.
However, the data are also consistent with the prediction of Drechsel 
et al.\cite{drechsel01} for a rapid variation at very small 
$Q^2$ and a departure from the GDH sum rule.
Future experiment\cite{chen01} at Hall-A will extend 
the $^3He$ measurement down to $Q^2=0.02$ GeV$^2$ in order to map out 
the low $Q^2$ behavior of the neutron generalized GDH integral.

\subsection{Quark-hadron duality}

The recent studies at JLab of the spin-averaged and spin-dependent 
structure functions at low $Q^2$ region have shed new light on the
subject of quark-hadron duality. Thirty years ago, Bloom and 
Gilman\cite{bloom70}
noticed that the structure functions obtained from deep-inelastic 
scattering experiments, where the substructures of the nucleon are
probed, are very similar to the averaged structure functions measured 
at lower energy, where effects of nucleon resonances dominate. 
This surprising similarity between the resonance electroproduction 
and the deep inelastic scattering suggests a common origin for these 
two phenomena, called local duality.

Recently, high precision data\cite{niculescu00} 
from JLab have verified the quark-hadron 
duality for spin-averaged scattering on proton and deuteron targets. 
For $Q^2$ as low as 0.5 GeV$^2$, the resonance data are within 10\%
of the DIS results. When the mean $F_2$ curve from the resonance data is
plotted as a function of the Nachtmann variable, 
$\xi = 2x/(1+\sqrt{1+4M^2x^2/Q^2})$, it resembles the $xF_3$ structure
function obtained in neutrino scattering experiments. Since $xF_3$
is a measure of the valence quark distributions, the similarity 
between $xF_3$ and the mean $F_2$ suggests that the 
$F_2$ structure function at low $Q^2$ originates 
from valence quarks only.

It is of much interest to extend the study of quark-hadron duality
to spin-dependent structure functions. 
In this Symposium, Griffioen\cite{griffioen01} presented the preliminary
result from CLAS showing that quark-hadron duality is also observed in
$g_1^p$. A comparison of $g_1^p(\xi)$
between the resonance data and the Gehrmann and 
Stirling\cite{gehrmann96} parameterization of the DIS data 
shows a good agreement for $\xi > 0.2$ (the valence quark region). 
At $\xi < 0.2$,
the resonance data appear to deviate from the DIS data. This trend
is reminiscent of what was observed for the spin-averaged structure 
functions\cite{niculescu00}. Additional information on the quark-hadron
duality is also expected from the JLab Hall-A measurement\cite{chen01} of 
$g_1^n(\xi)$ for neutrons.

\subsection{$g_2(x,Q^2)$}

Unlike the spin structure function $g_1$, which has a clear interpretation in
the quark-parton model (QPM), the $g_2$ structure function is sensitive to 
higher-twist quark-gluon correlation effect and is not readily interpreted in
QPM. The $g_2$ structure function probes both the transverse and the 
longitudinal parton distributions in the nucleons. Using the operator product
expansion (OPE) technique\cite{shuryak82,jaffe91} 
in QCD, $g_2$ can be expressed in terms of three 
components\cite{cortes92}: a leading twist-2 
part $g_2^{WW}(x,Q^2)$ originating from the same
set of operators that contribute to $g_1$, another twist-2 structure function
$h_T(x,Q^2)$ depicting quark transverse polarization, and a twist-3 part 
from quark-gluon interactions $\xi(x,Q^2)$,

\begin{equation}
g_2(x,Q^2) = g_2^{WW}(x,Q^2) - \int_{x}^{1} \frac{\partial}{\partial y}
(\frac{m}{M} h_T(y,Q^2) + \xi(y,Q^2))\frac{dy}{y},
\label{Eq:g2a}
\end{equation}

\noindent where $m$ and $M$ are quark and nucleon masses. 
The twist-2 contribution, $g_2^{WW}$, is related to $g_1$
via\cite{wandzura77} 

\begin{equation}
g_2^{WW}(x,Q^2) = -g_1(x,Q^2) + \int_{x}^{1} \frac{g_1(y,Q^2)}{y}
dy.
\label{Eq:g2c}
\end{equation}

The contribution from the transversity distribution $h_T(x,Q^2)$ 
is suppressed by the $m/M$ term and can be neglected. 
Hence, the difference between $g_2$ and $g_2^{WW}$, 
$\overline{g_2} = g_2 - g_2^{WW}$, will isolate the twist-3 contribution.

The moments of $g_1$ and $g_2$
can be derived from OPE:

\begin{eqnarray}
\int_{0}^{1} x^n g_1(x,Q^2) dx & = & 
\frac{a_n}{2},~~~~~~~~~~~~~~~~~~~~~~~  n=0,2,4,... \nonumber \\
\int_{0}^{1} x^n g_2(x,Q^2) dx & = & \frac{1}{2} \frac{n}{n+1} (d_n-a_n),  
~~~~~n=2,4,6,..., 
\label{Eq:g2b}
\end{eqnarray}

\noindent where $a_n$ and $d_n$ are the twist-2 and twist-3 matrix 
elements of the renormalized operators, respectively. 
The twist-3 matrix elements $d_n$ can then be evaluated using

\begin{equation}
d_n = \frac{2(n+1)}{n} \int_{0}^{1} x^n \overline{g_2} (x,Q^2) dx.
\label{Eq:g2d}
\end{equation}

\noindent A primary goal for measuring $g_2$ is therefore to determine
the twist-3 contribution which reflects quark-gluon correlation effects.
A series of SLAC experiments, E142, E143, E154, E155 and E155x, have
measured $g_2(x,Q^2)$ for protons and neutrons. In this Symposium, 
Bosted\cite{bosted01} reported the near-final results of $g_2^p$ and
$g_2^d$ from E155x. The $x$-dependence of all SLAC $g_2^p$ data is
reasonably well described by the twist-2 component $g_2^{WW}$.
However, the data also allow non-zero twist-3 contributions.
In particular, the twist-3 matrix elements, $d_2^p$ and $d_2^n$,
evaluated using Eq.~\ref{Eq:g2d}, are shown to be small but 
non-zero\cite{bosted01}. Meziani\cite{meziani01} discussed
in this Symposium a proposal to measure $g_2^n$ with high accuracy
at the proposed 12 GeV CEBAF upgrade using a polarized $^3He$
target and a large acceptance spectrometer. This could provide
a definitive result on the twist-3 content of the nucleon.

The new $g_2(x)$ data from E155x also allow evaluations of the integrals
$\int g_2(x) dx$ in order to check the Burkhardt-Cottingham sum rule.
In this Symposium, Bosted\cite{bosted01} reported that for $0.02 \le x
\le 0.8$ at $Q^2 = 5$ GeV$^2$, the integral was found to be $-0.034 \pm
0.008$ for proton and $-0.002 \pm 0.011$ for deuteron. The apparent
disagreement between the proton result and the Burkhardt-Cottingham
sum rule could be due to the contribution of the unmeasured small
$x$ region. Another sum rule by Efremov, Leader and Teryaev 
(ELT)\cite{efremov97}, derived with the assumption of isospin
symmetry of the sea-quark distributions, gives

\begin{equation}
\int_{0}^{1} x [g_1^p(x) + 2g_2^p(x) -g_1^n(x) -2 g_2^n(x)] dx
= 0.
\label{Eq:etd}
\end{equation}

\noindent The ELT sum rule is much less sensitive to the small-$x$ uncertainty due
to the factor $x$ in the integrand. The most recent value of this integral for 
the proton is found\cite{bosted01} to be $-0.009 \pm 0.008$ at 
$Q^2$ = 2.5 GeV$^2$, consistent
with the prediction of the sum rule. The recent JLab Hall-A data\cite{chen01} 
on $g_2^n$ could provide a further check of this sum rule for the neutron.

\section{Polarized Quark and Gluon Distributions}

\subsection{Polarized quark and antiquark distributions}

The $g_1(x,Q^2)$ data obtained in inclusive DIS have been analyzed\cite{SMC_98} 
in the framework of NLO QCD to extract information on $\Delta \Sigma (x)$, 
$\Delta q_{NS}^{p,n} (x)$, and $\Delta g(x)$, where $\Delta \Sigma$ and
$\Delta q_{NS}$ correspond to the flavor-singlet and flavor-nonsinglet
quark polarization, and $\Delta g$ is the gluon polarization.
The inclusive DIS data, however, do not allow a detailed flavor decomposition
of the nucleon spin. In particular, the contributions from valence and sea
quarks are not separated. To overcome this limitation, the SMC and HERMES
experiments have advocated the use of semi-inclusive deep inelastic scattering
(SIDIS), where the leading hadrons accompanying the DIS process are also detected.
The flavor of the struck quark is expected to be reflected by the flavor of the
produced hadrons. Hence the hadrons provide a ``tag" on the flavor of the
struck quark. This technique was first applied by the SMC\cite{smc96,smc98}
collaboration to determine $\Delta u_v$, $\Delta d_v$, and $\Delta \bar q
(\Delta \bar u = \Delta \bar d = \Delta \bar q)$. Later, the HERMES collaboration
used this method to measure the $\bar d - \bar u$ flavor asymmetry of the 
unpolarized nucleon sea and obtained\cite{hermes98} a result 
consistent with that from a completely 
different approach\cite{hawker98,peng98,garvey01} using the Drell-Yan process. 
This agreement suggests that the SIDIS is indeed a valid tool for studying 
flavor decomposition.

\begin{figure}[h]
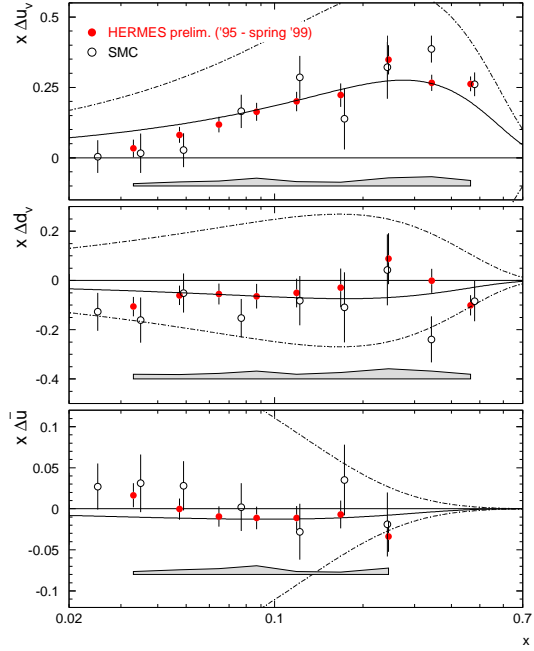

\PSFIG{peng4}{7cm}{0}
\caption{Preliminary result of the HERMES analysis of
SIDIS data$^{44}$. The data are presented at a common 
$Q^2$ of 2.5 GeV$^2$. The dashed dotted lines indicate
the positivity limit and the solid lines show parameterizations from
Gehrmann and Stirling.}
\label{fig:SIDIS1}
\end{figure}

Extensive SIDIS data have been collected\cite{hermes99} by 
the HERMES collaboration on
polarized gas targets of $H$, $D$, and $^3He$. 
Shibata\cite{shibata01} and Bernreuther\cite{bernreuther01} presented
in this Symposium the preliminary results of $\Delta u_v$, $\Delta d_v$,
and $\Delta \bar u$ based on an analysis of the HERMES data collected up to 
Spring '99. As shown in Fig~\ref{fig:SIDIS1}, the HERMES data have improved
accuracy over the SMC data. Both the SMC and the HERMES results are in very
good agreement with the parameterization of Gehrmann 
and Stirling\cite{gehrmann96}.
In order to increase the statistical significance, the constraint

\begin{equation}
\Delta \bar q/\bar q = \Delta u_s/u_s = \Delta \bar u/\bar u 
= \Delta d_s/d_s = \Delta \bar d/\bar d = \Delta \bar s/\bar s = \Delta s/s
\label{Eq:sidis1}
\end{equation}

\noindent has been imposed in the analysis shown in Fig.~\ref{fig:SIDIS1}.

\begin{figure}[h]
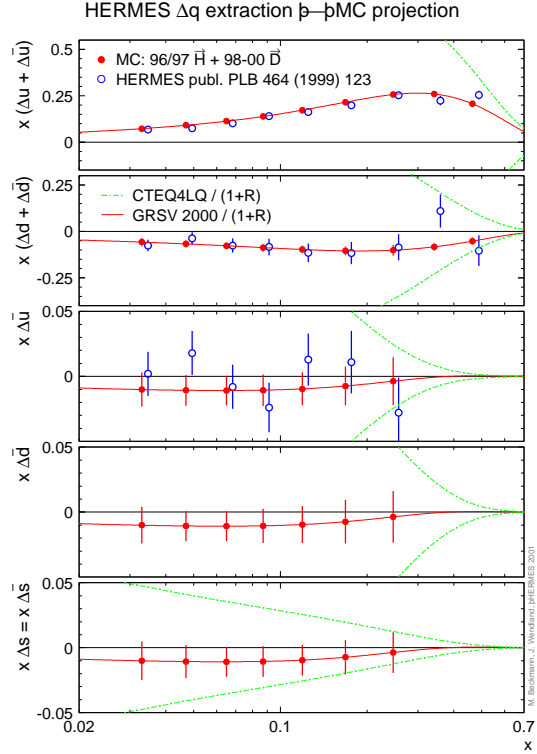

\PSFIG{peng5}{7cm}{0}
\caption{Monte Carlo prediction for the statistical precision on the extraction
of polarized quark distributions times Bjorken $x$ after analyzing all
HERMES data taken up to summer 2000$^{44}$.
The Monte Carlo points have been placed on the parameterization curve 
from GRSV2000. 
The dashed dotted lines represent the positivity limit based on the
unpolarized CTEQ4LQ parton distribution functions.}
\label{fig:SIDIS2}
\end{figure}

With the large sample of polarized $e^+ + d$ SIDIS data collected during
1999 and 2000 in conjunction with an operational RICH detector for $\pi/K/p$
identification, the HERMES collaboration is now analyzing their SIDIS data
without the contraint of Eq.~\ref{Eq:sidis1}. The identification of kaons
with the RICH detector would help isolating the $\Delta s$ component. 
The anticipated statistical accuracy for $\Delta u_v$, $\Delta d_v$,
$\Delta \bar u$, $\Delta \bar d$, and $\Delta s (= \Delta \bar s)$ 
in this 5-parameter analysis is
shown in Fig.~\ref{fig:SIDIS2}. The HERMES analysis could lead to
exciting first results on the flavor structure of the sea-quark
polarizations. Several interesting issues, such as the large flavor
asymmetry between the $\Delta \bar u$ and $\Delta \bar d$ 
predicted\cite{diakonov97}
in the chiral-quark-soliton model as well as the negative strange-quark
polarization expected from the analysis of $g_1$, could be addressed.
The result of the five-parameter
analysis is expected\cite{shibata01} to be available soon.

Another promising technique for measuring sea-quark polarization is
$W$-boson production\cite{bourrely93,bourrely94} at RHIC.
The longitudinal single-spin asymmetry for $W$ production in
$\vec p + p \to W^{\pm} + x$ can be written in leading order as

\begin{equation}
A_L^{W^+} = \frac{\Delta u(x_1)\bar d(x_2) - \Delta \bar d(x_1) u(x_2)}
{u(x_1)\bar d(x_2) + \bar d(x_1) u(x_2)},~~
A_L^{W^-} = \frac{\Delta d(x_1)\bar u(x_2) - \Delta \bar u(x_1) d(x_2)}
{d(x_1)\bar u(x_2) + \bar u(x_1) d(x_2)},
\label{Eq:wprod1}
\end{equation}

\noindent where $x_{1,2}$ are the Bjorken-$x$ of the colliding quarks 
and antiquarks. For $x_1<<x_2$, Eq.~\ref{Eq:wprod1} becomes

\begin{equation}
A_L^{W^+} \approx - \frac{\Delta \bar d(x_1)}{\bar d(x_1)},~~
A_L^{W^-} \approx - \frac{\Delta \bar u(x_1)}{\bar u(x_1)}, 
\label{Eq:wprod2}
\end{equation}

\noindent and $A_L$ gives a direct measure of sea-quark polarization.
For $x_1 >> x_2$, one obtains

\begin{equation}
A_L^{W^+} \approx - \frac{\Delta u(x_1)}{u(x_1)},~~
A_L^{W^-} \approx - \frac{\Delta d(x_1)}{d(x_1)}, 
\label{Eq:wprod3}
\end{equation}

\noindent and the valence quark polarization is probed. In this Symposium,
Kiryluk\cite{kiryluk01} discussed the plan to measure 
$W^{\pm} \to e^{\pm} + x$ at the STAR collaboration. The PHENIX collaboration
is capable of measuring the $W^{\pm} \to \mu^{\pm} + x$ decays as well,
as discussed by Saito\cite{saito01}. The RHIC $W$-production and
the PHENIX SIDIS measurements are clearly complementary tools for
determining polarized quark and antiquark distributions.

\subsection{Polarized gluon distribution}

Analysis of existing $g_1$ data showed that only $\sim$ 30\% of the
nucleon's spin is carried by quarks. This suggests that gluons could have
a large polarization. A NLO analysis of $g_1(x,Q^2)$ by the 
SMC collaboration\cite{SMC_98} 
showed that $\Delta G(x)$ is positive, albeit with a large uncertainty. 
Global fits\cite{gehrmann96} to existing spin-dependent structure functions can also
accommodate very different parameterizations of $\Delta G(x)$, again
showing that gluon polarization is poorly known. A direct and precise 
determination of $\Delta G(x)$ remains one of the most important goals of
spin physics. 

Although inclusive DIS does not directly probe the gluon distribution,
certain semi-inclusive DIS processes are sensitive to $\Delta G$.
The HERMES collaboration has reported a measurement\cite{hermes00} of $\Delta G/G$
at $x \approx 0.17$ using semi-inclusive hadron-pair production. Lepto-
and photo- open-charm production has also been proposed at CERN\cite{compass} and at
SLAC\cite{bosted01,bosted00} for $\Delta G/G$ measurements.

Polarized proton-proton collision at RHIC offers a great opportunity
for studying gluon polarization. In this Symposium, Kiryluk\cite{kiryluk01}
discussed the proposed double longitudinal spin asymmetry, $A_{LL}$,
measurement for $\vec p + \vec p \to \gamma + jet + x$ at STAR and 
Liu\cite{liu01} discussed the heavy-quark production at PHENIX.
Other processes sensitive to $\Delta G$ are the inclusive prompt photon
production, jet production, and high-$p_T$ hadron production. Although
no new experimental results on $\Delta G(x)$ were presented in this
Symposium, it is clear that a wealth of new results will be forthcoming.

\section{Transversity Distributions}

In addition to the unpolarized and polarized quark distributions, $q(x,Q^2)$
and $\Delta q(x,Q^2)$, a third quark distribution, called transversity, is the
remaining twist-2 distribution yet to be measured. 
This helicity-flip quark distribution,
$\delta q(x,Q^2)$, can be described in QPM as the net transverse 
polarization of quarks in a transversely polarized nucleon. The corresponding
structure function is given by

\begin{equation}
h_1(x,Q^2) = \frac{1}{2} \sum_{i} e_i^2 \delta q_i(x,Q^2).
\label{Eq:trans1}
\end{equation}

Due to the chiral-odd nature of the transversity distribution, it can not
be measured in inclusive DIS experiments. In order to measure $\Delta q(x,Q^2)$,
an additional chiral-odd object is required. For example, the double
spin asymmetry, $A_{TT}$, for Drell-Yan cross section 
in transversely polarized $p p$ collision, is sensitive to transversity
since $A_{TT} \sim \sum_{i} e_i^2 \delta q_i(x_1) \delta \bar q_i(x_2)$.
Such a measurement could be carried out at RHIC\cite{bunce00,saito01}, 
although the anticipated effect is small, on the order of $1-2$\%.

Several other methods for measuring transversity have been proposed for
semi-inclusive DIS. In particular, Collins suggested\cite{collins93} 
that a chiral-odd
fragmentation function in conjunction with the chiral-odd transversity
distribution would lead to an observable single-spin azimuthal asymmetry
in semi-inclusive pion production. An analysis of the jet structure in
$Z^\circ \to 2$ jets decay suggested that the Collins function has
a sizable magnitude\cite{efremov99}.

\begin{figure}[t]
\begin{center}
\epsfxsize=29pc
\epsfbox{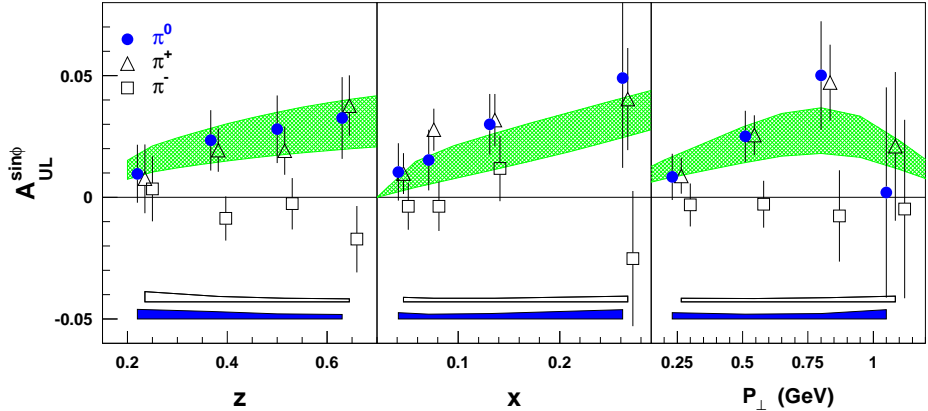}
\end{center}
\vspace*{-6mm}
\caption{Analyzing power in the $\sin \phi$ moment from 
HERMES$^{57,58}$. Error bars
include the statistical uncertainty only. The filled and open bands at the
bottom of the panels represent the systematic uncertainties for neutral and
charged pions, respectively. The shaded areas show a range of
predictions of a model calculation applied to the case of $\pi^\circ$
electro-production$^{59,60}$.}
\label{fig:trans1}
\vspace*{-4mm}
\end{figure}

As reported by Schnell\cite{schnell01} in this Symposium, the HERMES 
collaboration has recently measured\cite{hermes01} 
single-spin azimuthal asymmetry
for charged and neutral pion electroproduction. Using
unpolarized positron beam on a longitudinally polarized hydrogen
target, the cross section was found to have a sin$\phi$ dependence, where
$\phi$ is the azimuthal angle between the pion and the $(e, e^\prime)$
scattering plane.
This Single-Spin-Asymmetries (SSA) can be expressed as the analyzing power
in the sin$\phi$ moment, and the result is shown in Fig.~\ref{fig:trans1}
for $\pi^+$, $\pi^-$, and $\pi^\circ$ as a function of the pion fractional 
energy $z$, the Bjorken $x$, and the pion transverse momentum $P_{\bot}$. 
\noindent The sin$\phi$ moment for an unpolarized (U) positron scattered off 
a longitudinally (L) polarized target contains two main contributions

\begin{eqnarray}
\langle sin \phi \rangle & \alpha & S_L \frac{2 (2-y)}{Q\sqrt{1-y}}
\sum_{q} e_q^2 x h_L^q(x) H_1^{\bot,q}(z) \nonumber \\
& + & S_T (1-y) \sum_{q} e_q^2 x h_1^q(x) H_1^{\bot,q}(z),
\label{Eq:ssa1}
\end{eqnarray}
\noindent where $S_L$ and $S_T$ are the longitudinal and transverse components
of the target spin orientation with respect to the virtual photon direction.
For the HERMES experiment with a longitudinally polarized target, the
transverse component is nonzero with a mean value of $S_T \approx 0.15$.
The observed azimuthal asymmetry could be a combined effect of the
$h_1$ transversity and the twist-3 $h_L$ distribution. Figure~\ref{fig:trans1}
shows that a model calculation\cite{Oganessian98,DeSanctis00} 
reproduces the $z$, $x$, and $P_{\bot}$ dependences
of the $\pi^\circ$ asymmetry quite well. The striking difference between the $\pi^+$
and $\pi^-$ analyzing power suggests that the Collins fragmentation function is
sizable only when the flavor of the truck quark is present in the final hadron.

If the azimuthal asymmetry observed by HERMES is indeed caused by the
$h_1$ transversity, a much larger asymmetry is expected for a transversely
polarized target. An earlier SMC measurement had limited statistics and
was inconclusive\cite{smc99}. The HERMES Collaboration plans to measure\cite{schnell01}
the shape of $\delta u(x)$ (and $H_1^{\bot,u}(z)$) using a transversely polarized
proton target in 2002-03. A proposal to measure $\delta d(x)$ using 
a transversely polarized
deuterium target has also been discussed\cite{korotkov01}.

\section{Generalized Parton Distributions and DVCS}

There has been intense theoretical and experimental activities in recent years
on the subject of Generalized Parton Distribution (GPD). In the Bjorken scaling
regime, exclusive leptoproduction reactions can be factorized into a hard-scattering
part calculable in QCD, and a non-perturbative part parameterized by the GPDs.
The GPD takes into account dynamical correlations between partons with different
momenta. In addition to the dependence on $Q^2$ and $x$, the GPD also depends on
two more parameters, the skewedness $\xi$ and the momentum transfer to the 
baryon, $t$. Of particular interest is the connection between GPD and the nucleon's
orbital angular momentum\cite{ji97}.

The deeply virtual Compton scattering (DVCS), in which an energetic photon is
produced in the reaction $e p \to e p \gamma$, is most suitable for studying GPD.
Unlike the exclusive meson productions, DVCS avoids the complication associated with
mesons in the final state and can be cleanly interpreted in terms of GPDs.
An important experimental challenge, however, is to separate the relatively
rare DVCS events from the abundant electromagnetic Bethe-Heitler (BH) background.
Significant progress has been made recently, and several experiments at HERA and
JLab have reported observation of the DVCS events. From the collision
of 800 GeV protons with 27.5 GeV positrons, both the ZEUS\cite{saull00} 
and the H1\cite{adloff01} collaborations at DESY observed an excess of 
$e^+ + p \to e^+ + \gamma + p$ events in a kinematic region where 
the BH cross section is largely suppressed. The excess events were
attributed to the DVCS process and the H1 collaboration 
further determined\cite{adloff01} the DVCS cross section over the 
kinematic range $2 < Q^2 < 20$ GeV$^2$, $30 < W < 120$ GeV, 
and $|t| < 1$ GeV$^2$. 
(The $x$ range covered is roughly $0.00035 < x < 0.0035$ for $W = 75$ GeV.)

\begin{figure}[h]
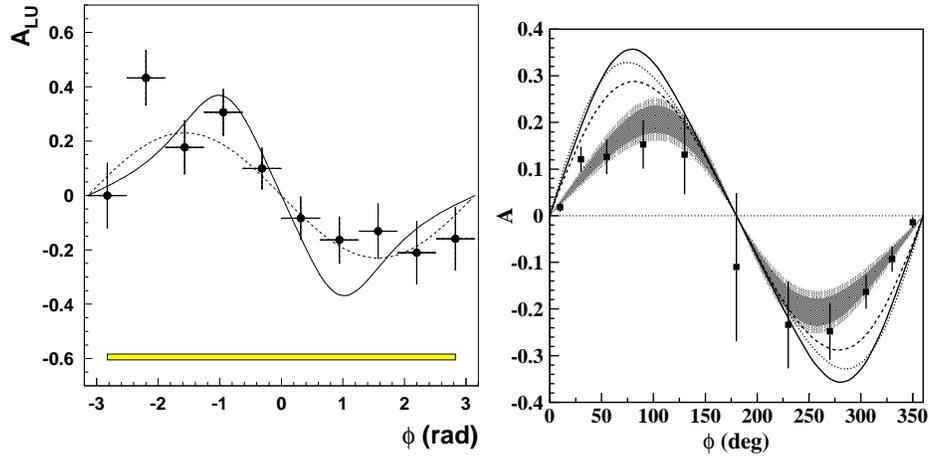

\PSFIG{peng7}{13cm}{0}
\caption{Left panel: HERMES data$^{66}$ 
on the beam-spin asymmetry for hard
exclusive electroproduction of photons as a function
of the azimuthal angle $\phi$. Right panel: CLAS 
data$^{67}$ on the
beam-spin asymmetry resulting from the DVCS-BH interference.}
\label{fig:dvcs1}
\end{figure}

At lower c.m. energies, the HERMES\cite{hermes01a} and the 
CLAS\cite{stepanyan01} collaborations
observed the interference between the DVCS and the BH processes,
which manifests itself as a pronounced azimuthal asymmetry correlated with
the beam helicity. In this Symposium, Bianchi\cite{bianchi01} reported
the HERMES measurement\cite{hermes01a} 
shown in Fig.~\ref{fig:dvcs1}. The HERMES result
is in nice agreement with the CLAS result\cite{stepanyan01}, also shown 
in Fig.~\ref{fig:dvcs1}. Note that there exists several differences between
these two measurements. First, the beam energy for the CLAS experiment
(4.25 GeV) is lower than for the HERMES experiment (27.6 GeV). Second,
the CLAS experiment detected the electron and proton, while the HERMES
experiment measured the positron and the photon in the final state.
Finally, a polarized electron beam was used for CLAS instead of the polarized
positron beam for HERMES. The qualitative agreement between these two experiments
is reassuring. It is interesting to note that the CLAS data showed an opposite
sign for the azimuthal asymmetry relative to the HERMES data (in Fig.~\ref{fig:dvcs1}
the offsets of the azimuthal $\phi$ angle are different for 
the two experiments). This is
to be expected since the interference term is proportional to the sign of the
lepton charge. 

As discussed by Bianchi\cite{bianchi01} in this Symposium, another observable
sensitive to the interference between the DVCS and the BH processes
is the azimuthal asymmetry between unpolarized $e^+$ and $e^-$ beams. In contrast
to the Beam Spin Asymmetry (BSA) which is sensitive to the imaginary part of
the DVCS amplitudes, the Beam Charge Asymmetry (BCA) is probing the real part
of the DVCS amplitudes\cite{diehl97}. Analysis of the HERMES $e^-$ data 
in 98-99 and the
$e^+$ data in 99-00 is underway, and the first result on the BCA is expected
soon\cite{bianchi01}.

\section{Other Related Topics}

\subsection{$\Lambda$ polarization}

Schnell\cite{schnell01} described the recent progress at HERMES on
semi-inclusive $\Lambda$ production. The HERMES collaboration has measured
the longitudinal spin transfer from the polarized electron beam to
the $\Lambda$ particle in the final state, as well as the transverse $\Lambda$
polarization using unpolarized beam. The main interest for the longitudinal
spin transfer measurement is to deduce information on the polarized quark
distribution in the $\Lambda$. Assuming $u$-quark dominance as well as helicity
conservation in the $u \to \Lambda$ fragmentation process, the HERMES measurement
of the longitudinal spin transfer can reveal the role of $u$ quark for $\Lambda$
spin. Earlier measurements\cite{aleph96,opal98}
of $\Lambda$ polarization from $Z$ decays are
sensitive to $s$ quark polarization in the $\Lambda$. The NOMAD collaboration has
also measured\cite{nomad00,nomad01}
$\Lambda$ and $\bar \Lambda$ polarization in neutrino DIS.

The new HERMES measurement\cite{schnell01} covers a
kinematic range (roughly $0.2 < z < 0.8$ and
$0 < x_F < 0.8$) much broader than previous DIS
experiments\cite{adams00,airapetian01}.
Preliminary result on the average spin transfer in this kinematic
region is $0.04 \pm 0.09$. The $x_F$ dependence is consistent with
the previous NOMAD result\cite{nomad00}. The predictions of several
model calculations\cite{deflorian98,ma00} for a rising
spin transfer with increasing $z$
are not observed in the data, although the statistical accuracy
is limited. A new detector was installed at HERMES to enhance the
acceptance of $\Lambda$ detection for future measurements\cite{schnell01}.

\subsection{Exclusive electroproduction of mesons}

New results on exclusive meson production from HERMES and CLAS collaborations
have been presented in this Symposium. The current interest in hard exclusive processes
is largely due to the fact that QCD factorization was proved to be valid
for exclusive meson production with longitudinal virtual photons\cite{collins97}.
Such factorization allowed new means to extract the unpolarized and polarized
GPD. In particular, unpolarized GPDs can be measured
with exclusive vector meson production, while polarized GDPs can be probed via
exclusive pseudoscalar meson production.

Bianchi\cite{bianchi01} presented the
preliminary HERMES results on the longitudinal component of the 
exclusive $\rho$ and $\phi$ meson production cross sections. The data are in
good agreement with a calculation\cite{vanderhaeghen99} based on GPD.
For exclusive pseudoscalar meson production, the HERMES collaboration measured
the single spin azimuthal asymmetry in the reaction $e^+ + \vec p \to e^{\prime +}
+ n + \pi^+$. The asymmetry is found\cite{bianchi01,airapetian02} to be very 
large ($-0.18 \pm 0.05 \pm 0.02$) and has a sign opposite to that of inclusive 
$\pi^+$ production~\cite{hermes00a}.

The CLAS collaboration recently reported\cite{griffioen01,clas02} 
the first measurement of double spin asymmetry
in the $\vec e + \vec p \to e^\prime + \pi^+ + n$ reaction.
This observable is sensitive to the contributions from various
resonances, and the data indicate the dominance of the helicity-1/2 contribution
in the second and third resonance regions.

\subsection{$A_1^n$ at large $x$}

A new measurement\cite{chen01} of the $A_1^n$ 
spin asymmetry has been carried out at
Hall-A in JLab using polarized $^3He$ target. The value of 
$A_1^n$ as $x \to 1$ is sensitive to the
underlying models describing valence quark dynamics in the nucleon.
Isospin and SU(6) symmetries predict $A_1^n \to 1$ as $x \to 1$, while
pQCD approaches\cite{farrar75,brodsky95} predict 
$A_1^n \to 0$ as $x$ approaches 1. Preliminary JLab result\cite{chen01} 
for $A_1^n$ in the $0.33 < x < 0.61$ range show that $A_1^n$ turns positive
for large $x$. A definitive measurement of $A_1^n$ with high accuracy at 
very large $x$
has been considered for the 12 GeV CEBAF upgrade\cite{meziani01}.

\section{Summary and Outlook}

There has been an enormous progress in various areas of spin physics experiments
since the last Circum-Pan-Pacific Spin Symposium in 1999. An incomplete list
of the major progress would include the following:

\begin{itemize}

\item First JLab measurements on spin structure functions have provided
new information on $g_1^p, g_1^n, g_2^n, \Gamma_1^p, \Gamma_1^n$ and the
generalized GDH sum at the resonance region.

\item Observation of the azimuthal Single Spin Asymmetry at HERMES in
semi-inclusive pion production holds a great promise for measuring
transversity in the near future.

\item Observation of the DVCS process at HERMES and JLab has generated 
much interest and could develop into an extensive program for measuring
Generalized Parton Distributions.

\item Various spin observables in exclusive meson electroproductions
are being measured at HERMES and JLab.

\item Commissioning of the RHIC-spin has been successfully carried out\cite{bunce01}.

\end{itemize}

Many new results are anticipated in the near future:

\begin{itemize}

\item First results from RHIC-spin.

\item First results on $\Delta \bar u, \Delta \bar d$ and $\Delta s (= \Delta \bar s$)
from the 5-parameter analysis of the HERMES SIDIS data.

\item First measurement of the transversity ($\delta u$) at HERMES using
transversely polarized targets.

\item Precise data on DVCS from HERMES and JLab.

\item $\Delta G(x)$ from SLAC, COMPASS, HERMES, and RHIC-spin. 

\item Additional low-$Q^2$ data from JLab.

\item And much more ....
\end{itemize}

\section*{Acknowledgements}

I would like to thank Professor Bo-Qiang Ma for inviting me to attend
this very interesting and successful symposium.


\begin{thebibliography}{0}
\bibitem{filippone01}
B. W. Filippone and X. Ji,
hep-ph/0101224.
\bibitem{bunce00}
G. Bunce, N. Saito, J. Soffer, and W. Vogelsang,
{\it Ann. Rev. Nucl. Part. Sci.} {\bf 50}, 525 (2000).
\bibitem{jaffe01}
R. L. Jaffe, hep-ph/0101224.
\bibitem{anselmino01}
M. Anselmino, hep-ph/0107093.
\bibitem{jakob01}
N. Bianchi and R. Jakob, hep-ph/0108078.
\bibitem{makins00}
N. C. R. Makins, Talk presented at DIS2000.
\bibitem{E155_00}
The E155 Collaboration, P. L. Anthony {\it et al.}, {\it Phys. Lett.}
{\bf B493}, 19 (2000).
\bibitem{SMC_98}
The SMC Collaboration, B. Adeva {\it et al.}, {\it Phys. Rev.}
{\bf D58}, 112001 (1998).
\bibitem{SMC_98a}
The SMC Collaboration, B. Adeva {\it et al.}, {\it Phys. Rev.}
{\bf D58}, 112002 (1998).
\bibitem{ellis74}
J. Ellis and R. Jaffe, {\it Phys. Rev.}
{\bf D9}, 1444 (1974); {\bf D10}, 1669 (1974).
\bibitem{bjorken66}
J. D. Bjorken, {\it Phys. Rev.} {\bf 148}, 1467 (1966); {\bf D1}, 1376 (1970).
\bibitem{close94}
F. E. Close and R. G. Roberts, {\it Phys. Lett.} {\bf B336}, 257 (1994).
\bibitem{kwiecinski99}
J. Kwiecinski and B. Ziaja {\it Phys. Rev.} {\bf D60}, 054004 (1999).
\bibitem{hughes99}
V. W. Hughes and A. Deshpande, {\it Nucl. Phys. Proc. Suppl.}
{\bf 79}, 579 (1999), hep-ex/9906006.
\bibitem{gerasimov66}
S. B. Gerasimov, {\it Sov. J. Nucl. Phys.}
{\bf 2}, 430 (1966).
\bibitem{drell66}
S. D. Drell and A. C. Hearn, {\it Phys. Rev. Lett.}
{\bf 16}, 430 (1966).
\bibitem{ahrens01}
The GDH and A2 Collaborations J. Ahrens {\it et al.},
{\it Phys. Rev. Letts.} {\bf 87}, 022003 (2001), hep-ex/0105089.
\bibitem{drechsel99}
D. Drechsel {\it et al.}, {\it Nucl. Phys.} {\bf A645}, 145 (1999).
\bibitem{bianchi99}
N. Bianchi and T. Thomas, {\it Phys. Lett.} {\bf B450}, 439 (1999).
\bibitem{burkhardt70}
H. Burkhardt and W. N. Cottingham, {\it Ann. Phys. (N.Y.)} {\bf 56},
453 (1970).
\bibitem{griffioen01}                        
K. Griffioen, Talk presented at this Symposium (2001).
\bibitem{soffer93}                                    
J. Soffer and O. V. Teryaev, {\it Phys. Rev. Lett.} {\bf 70}, 3371 (1993).
\bibitem{burkert92}                                                       
V. Burkert and B. Ioffe, {\it Phys. Lett.} {\bf B296}, 223 (1992);
{\it J. Exp. Theo. Phys.} {\bf 78}, 619 (1994).                   
\bibitem{chen01}                               
J. P. Chen, Talk presented at this Symposium (2001).
\bibitem{drechsel01}                                
D. Drechsel {\it et al.}, {\it Phys. Rev.} {\bf D63}, 114010 (2001).
\bibitem{bloom70}                                                   
E. D. Bloom and F. J. Gilman, {\it Phys. Rev. Lett.} {\bf 25}, 1140 (1970);
{\it Phys. Rev.} {\bf D4}, 2901 (1971).                                    
\bibitem{niculescu00}                  
I. Niculescu {\it et al.}, {\it Phys. Rev. Lett.} {\bf 85}, 1182, 1186 (2000).
\bibitem{gehrmann96}                                                          
T. Gehrmann and W. J. Stirling, {\it Phys. Rev.} {\bf D53}, 6100 (1996).
\bibitem{shuryak82}                                                     
E. Shuryak and A. Vainshtein, {\it Nucl. Phys.} {\bf B201}, 141 (1982).
\bibitem{jaffe91}                                                      
R. Jaffe and X. Ji, {\it Phys. Rev.} {\bf D43}, 724 (1991).
\bibitem{cortes92}                                         
J. L. Cortes, B. Pire and J. P. Ralston, {\it Z. Phys.} {\bf C55}, 409 (1992).
\bibitem{wandzura77}                                                          
S. Wandzura and F. Wilczek, {\it Phys. Lett.} {\bf B72}, 195 (1977).
\bibitem{bosted01}                               
P. Bosted, Talk presented at this Symposium (2001).
\bibitem{meziani01}                               
Z. Meziani, Talk presented at this Symposium (2001).
\bibitem{efremov97}
A. V. Efremov, O. V. Teryaev and E. Leader, {\it Phys. Rev.} {\bf D55}, 4307 
(1997).
\bibitem{smc96}
The SMC Collaboration, B. Adeva {\it et al.}, {\it Phys. Lett.}
{\bf B369}, 93 (1996).
\bibitem{smc98}
The SMC Collaboration, B. Adeva {\it et al.}, {\it Phys. Lett.}
{\bf B420}, 180 (1998).
\bibitem{hermes98}
The HERMES Collaboration, K. Ackerstaff {\it et al.}, {\it Phys. Rev. Lett.}
{\bf 81}, 5519 (1998).
\bibitem{hawker98}
The E866 Collaboration, E. A. Hawker {\it et al.}, {\it Phys. Rev. Lett.}
{\bf 80}, 3715 (1998).
\bibitem{peng98}
The E866 Collaboration, J. C. Peng {\it et al.}, {\it Phys. Rev.}
{\bf D58}, 092004 (1998).
\bibitem{garvey01}
G. T. Garvey and J. C. Peng, {\it Prog. Part. Nucl. Phys.}
{\bf 47}, 203 (2001), nucl-ex/0109010.
\bibitem{hermes99}
The HERMES Collaboration, K. Ackerstaff {\it et al.}, {\it Phys. Lett.}
{\bf B464}, 123 (1999).
\bibitem{shibata01}                               
T.-A. Shibata, Talk presented at this Symposium (2001).
\bibitem{bernreuther01}                               
S. Bernreuther, Talk presented at this Symposium (2001).
\bibitem{diakonov97}
D. I. Diakonov {\it et al.}, {\it Phys. Rev.} {\bf D56}, 4069
(1997).
\bibitem{bourrely93}
C. Bourrely and J. Soffer, {\it Phys. Lett.}
{\bf B314}, 132 (1993).
\bibitem{bourrely94}
C. Bourrely and J. Soffer, {\it Nucl. Phys.}
{\bf B423}, 329 (1994).
\bibitem{kiryluk01}                               
J. Kiryluk, Talk presented at this Symposium (2001).
\bibitem{saito01}
N. Saito, Talk presented at this Symposium (2001).
\bibitem{hermes00}
The HERMES Collaboration, A. Airapetian {\it et al.}, {\it Phys. Rev. Lett.}
{\bf 84}, 2584 (2000).
\bibitem{compass}
COMPASS proposal, CERN/SPSLC-96-14 (March, 1996).
\bibitem{bosted00}
http://www.slac.stanford.edu/exp/e161.
\bibitem{liu01}
M. Liu, Talk presented at this Symposium (2001).
\bibitem{collins93}
J. Collins, {\it Nucl. Phys.}
{\bf B396}, 161 (1993); {\it Nucl. Phys.} {\bf B420}, 565 (1994).
\bibitem{efremov99}
A. V. Efremov, O. G. Smirnova and L. G. Tkachev, {\it Nucl. Phys.
Proc. Suppl.} {\bf 74}, 49 (1999).
\bibitem{schnell01}
G. Schnell, Talk presented at this Symposium (2001).
\bibitem{hermes00a}
The HERMES Collaboration, A. Airapetian {\it et al.}, {\it Phys. Rev. Lett.}
{\bf 84}, 4047 (2000).
\bibitem{hermes01}
The HERMES Collaboration, A. Airapetian {\it et al.}, {\it Phys. Rev.}
{\bf D64}, 097101 (2001).
\bibitem{Oganessian98}
K.~A.~Oganessian, H.~R.~Avakian, N.~Bianchi and A.~M.~Kotzinian,
hep-ph/9808368.
\bibitem{DeSanctis00}
E.~De Sanctis, W.-D.~Nowak and K.~A.~Oganessian,
{\it Phys. Lett.} {\bf B483}, 69 (2000).
\bibitem{smc99}
A. Bravar for the SMC Collaboration,
{\it Nucl. Phys. Proc. Suppl.} {\bf 79}, 520 (1999).
\bibitem{korotkov01}
V. A. Korotkov, W. -D. Nowak and K. A. Oganessian, 
{\it Eur. Phys. J.} {\bf C18}, 639 (2001).
\bibitem{ji97}
X. Ji, {\it Phys. Rev. Lett.} {\bf 78}, 610 (1997);
{\it Phys. Rev.} {\bf D55}, 7114 (1997).
\bibitem{saull00}
The ZEUS Collaboration, P. R. B. Saull, Proceedings of the International Europhysics
Conference on High-Energy Physics, Tampere, Finland, 420-422 (1999),
hep-ex/0003030.
\bibitem{adloff01}
The H1 Collaboration, C. Adloff {\it et al.}, {\it Phys. Lett.}
{\bf B517}, 47 (2001).
\bibitem{hermes01a}
The HERMES Collaboration, A. Airapetian {\it et al.}, {\it Phys. Rev. Lett.}
{\bf 87}, 182001 (2001).
\bibitem{stepanyan01}
The CLAS Collaboration, S. Stepanyan {\it et al.}, {\it Phys. Rev. Lett.}
{\bf 87}, 182002 (2001).
\bibitem{bianchi01}
N. Bianchi, Talk presented at this Symposium (2001).
\bibitem{diehl97}
M. Diehl {\it et al.}, {\it Phys. Lett.} {\bf B411}, 193 (1997).
\bibitem{aleph96}
The ALEPH Collaboration, D. Buskulic {\it et al.}, {\it Phys. Lett.}
{\bf B374}, 319 (1996).
\bibitem{opal98}
The OPAL Collaboration, K. Ackerstaff {\it et al.}, {\it Eur. Phys. J.}
{\bf C2}, 49 (1998).
\bibitem{nomad00}
The NOMAD Collaboration, P. Astier {\it et al.}, {\it Nucl. Phys.}
{\bf B588}, 3 (2000).
\bibitem{nomad01}
The NOMAD Collaboration, P. Astier {\it et al.}, {\it Nucl. Phys.}
{\bf B605}, 3 (2001).
\bibitem{adams00}
The E665 Collaboration, M. R. Adams {\it et al.}, {\it Eur. Phys. J.}
{\bf C17}, 263 (2000).
\bibitem{airapetian01}
The HERMES Collaboration, A. Airapetian {\it et al.}, {\it Phys. Rev.}
{\bf D64}, 112005 (2001).
\bibitem{deflorian98}
D. de Florian, M. Stratmann and W. Vogelsang, {\it Phys. Rev.}
{\bf D57}, 5811 (1998).
\bibitem{ma00}
B. Q. Ma, I. Schmidt and J. J. Yang,
{\it Phys. Lett.} {\bf B477}, 107 (2000).
\bibitem{collins97}
J. C. Collins {\it et al.}, {\it Phys. Rev.}
{\bf D56}, 2982 (1997).
\bibitem{vanderhaeghen99}
M. Vanderhaeghen {\it et al.}, {\it Phys. Rev.}
{\bf D60}, 094017 (1999).
\bibitem{airapetian02}
The HERMES Collaboration, A. Airapetian {\it et al.}, hep-ex/0112022.
\bibitem{clas02}
The CLAS Collaboration, R. De Vita {\it et al.}, {\it Phys. Rev. Lett.} {\bf 88},
082001 (2002).
\bibitem{farrar75}
G. R. Farrar and D. R. Jackson, {\it Phys. Rev. Lett.} {\bf 35}, 1416 (1975).
\bibitem{brodsky95}
S. Brodsky, M. Burkardt and I. Schmidt, {\it Nucl. Phys.} {\bf B441}, 197 (1995).
\bibitem{bunce01}
G. Bunce, Talk presented at this Symposium (2001).

\end{thebibliography}
\end{document}